\documentclass[twocolumn]{jpsj3}
\usepackage{graphicx}
\usepackage{color}
\usepackage{bm}

\title{%
Zero-Energy State Localized near an Arbitrary Edge
in Quadrupole Topological Insulators
}

\author{%
Yositake Takane
}

\inst{%
Department of Quantum Matter, Graduate School of Advanced Sciences of Matter,\\
Hiroshima University, Higashihiroshima, Hiroshima 739-8530, Japan
}

\recdate{ \hspace{50mm} }

\abst{%
A two-dimensional quadrupole topological insulator on a square lattice
is a typical example of a higher-order topological insulator.
It hosts an edge state localized near each of its $90^{\circ}$ corners
at an energy $E$ inside the band gap,
where $E$ is set equal to zero for simplicity.
Although the appearance of an edge state has been shown in simple systems
with only $90^{\circ}$ corners, it is uncertain whether
a similar localized state can appear at $E = 0$ near a complicated edge
consisting of multiple $90^{\circ}$ and $270^{\circ}$ corners.
Here, we present a numerical method to determine the wavefunction of
a zero-energy state localized near an arbitrary edge.
This method enables us to show that one localized state appears at $E = 0$
if the edge consists of an odd number of corners.
In contrast, the energy of localized states inevitably deviates from $E = 0$
if the edge includes an even number of corners.
}

\begin{document}
\sloppy
\maketitle

\section{Introduction}

A two-dimensional topological insulator (i.e., quantum spin Hall insulator)
hosts one-dimensional helical states
at its edge,~\cite{kane,bernevig,qi,murakami,konig}
whereas a three-dimensional topological insulator hosts two-dimensional helical
states on its surface.~\cite{fu,moore,roy,ando-exp}
That is, a $d$-dimensional topological insulator hosts
$(d-1)$-dimensional helical states at its boundary.
These helical states appear as midgap states
inside the energy gap of bulk states.

Recently, higher-order topological insulators have been proposed in
Refs.~\citen{benalcazar1} and \citen{benalcazar2} and have attracted
considerable attention.~\cite{liu,langbehn,song,hashimoto,khalaf,ezawa1,
ezawa2,fukui1,matsugatani,hayashi,trifunovic,araki,fukui2,peterson,
serra-garcia,schindler}
A two-dimensional second-order topological insulator hosts zero-dimensional
states at its corners, whereas a three-dimensional second-order
(third-order) topological insulator hosts zero-dimensional
(one-dimensional) states at its corners (edges).
That is, a $d$-dimensional $\mathcal{D}$th-order topological insulator
hosts $(d-\mathcal{D})$-dimensional states at its boundary,
where $2 \le \mathcal{D} \le d$.
These states also appear as midgap states.

We focus on a two-dimensional second-order topological insulator
on a square lattice.
This is referred to as a quadrupole topological insulator
as it hosts zero-dimensional states localized near four corners
in a rectangular system.
Hereafter, a zero-dimensional state is referred to as a corner state
and its energy $E$ is set equal to zero (i.e., $E = 0$).~\cite{comment}
Previously, the appearance of corner states has been shown
in simple systems, such as the rectangular system
with four $90^{\circ}$ corners.~\cite{benalcazar1,benalcazar2}
However, it is uncertain whether a similar localized state can appear
at $E = 0$ near a complicated edge
consisting of multiple $90^{\circ}$ and $270^{\circ}$ corners.

In this paper, we present a numerical method of determining
the wavefunction of a zero-energy state localized near an arbitrary edge
consisting of multiple $90^{\circ}$ and $270^{\circ}$ corners.
As a byproduct, we find that a zero-energy state appears only when
the number of corners is an odd integer.
In other words, the energy of localized states inevitably deviates from
$E = 0$ if the number of corners is an even integer.
In the next section, we introduce a tight-binding model of
quadrupole topological insulators on a square lattice,
which possesses chiral symmetry.
In Sect.~3, we present a numerical method of determining the wavefunction of
a zero-energy state in semi-infinite systems and apply it
to the simple case with one $90^{\circ}$ or $270^{\circ}$ corner.
In Sect.~4, we determine the wavefunction of a zero-energy state
in semi-infinite systems
with multiple $90^{\circ}$ and $270^{\circ}$ corners.
The last section is devoted to a summary and discussion.
It is pointed out that our argument relies on the chiral symmetry.

\section{Model}

\begin{figure}[btp]
\begin{tabular}{cc}
\begin{minipage}{0.5\hsize}
\begin{center}
\hspace{-10mm}
\includegraphics[height=3.6cm]{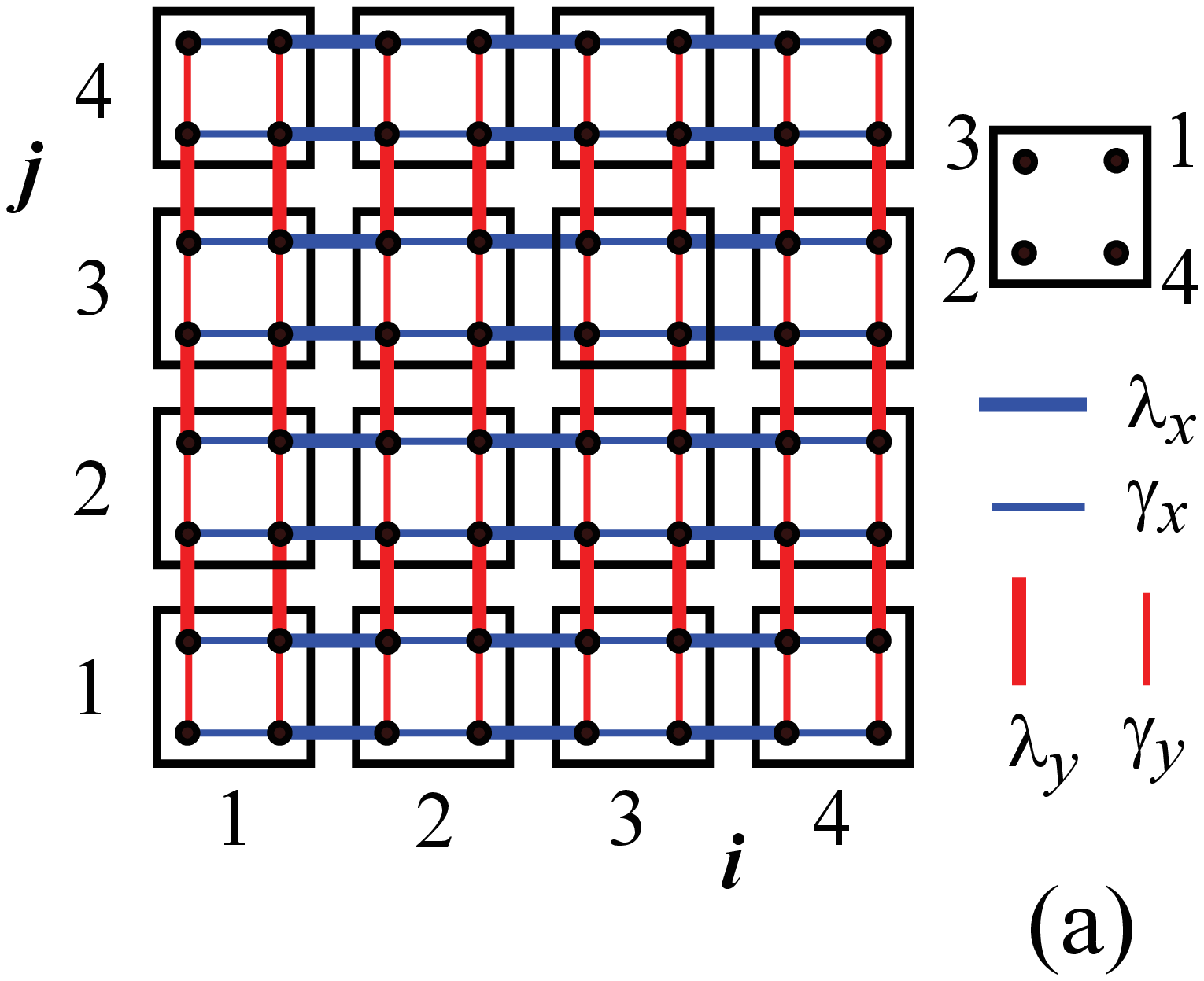}
\end{center}
\end{minipage}
\begin{minipage}{0.5\hsize}
\begin{center}
\hspace{-5mm}
\includegraphics[height=3.6cm]{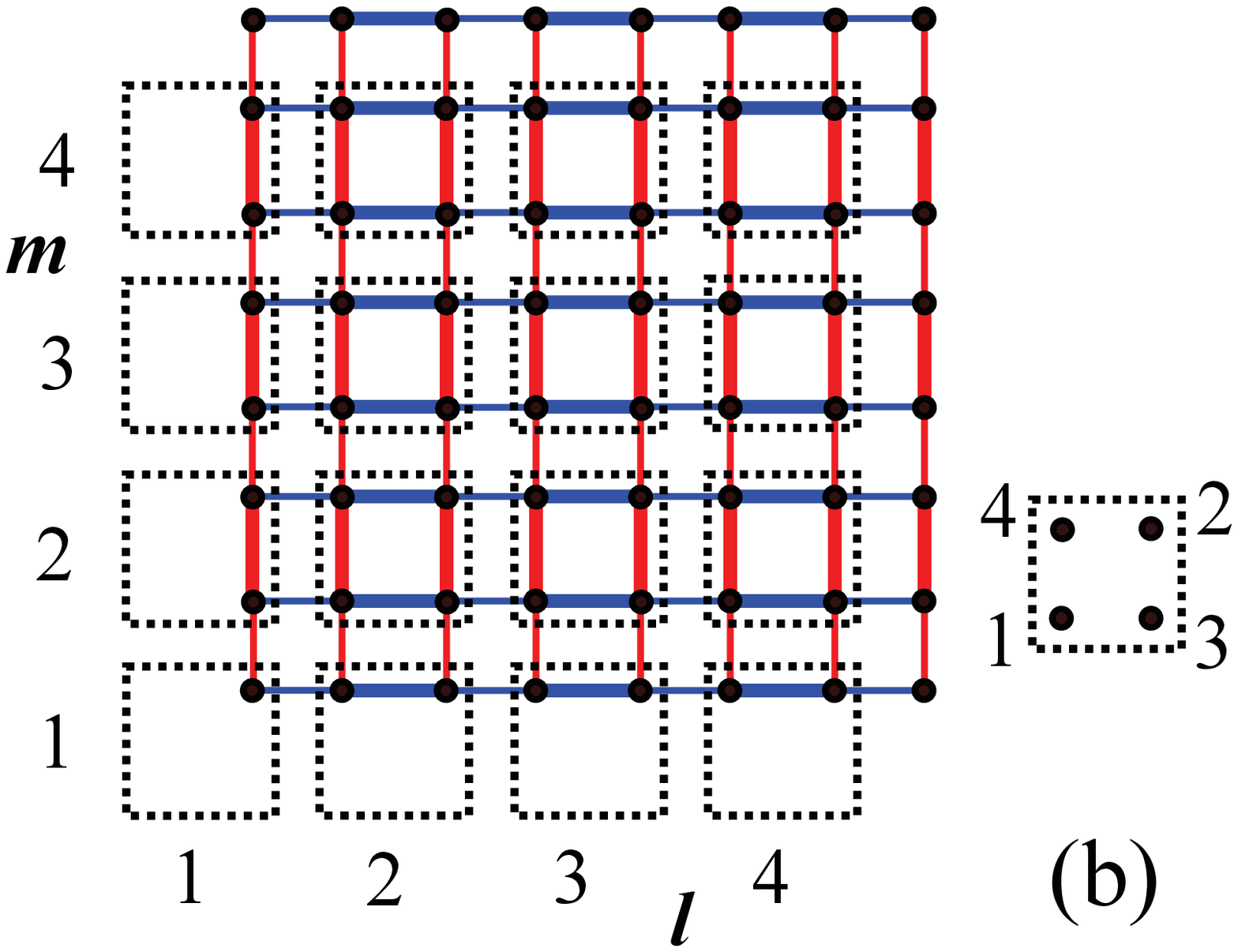}
\end{center}
\end{minipage}
\end{tabular}
\caption{
(Color online) Model system on a square lattice.
Thick horizontal (vertical) lines represent $\lambda_{x}$ ($\lambda_{y}$)
and thin horizontal (vertical) lines represent $\gamma_{x}$ ($\gamma_{y}$).
(a) Each solid square represents a unit cell, and
(b) each dotted square represents a dual cell.
}
\end{figure}
We introduce a tight-binding model for quadrupole topological insulators
on a square lattice with lattice constant $a$, where the unit cell
consists of four sites numbered by $1$, $2$, $3$, and $4$
as shown in Fig.~1(a).
Each unit cell is characterized by indices $i$ and $j$ respectively
specifying its location in the $x$- and $y$-directions.
The four-component state vector for the $(i,j)$th unit cell is expressed as
\begin{align}
  |i,j \rangle
  =  \bigl\{ |i,j \rangle_{1}, |i,j \rangle_{2},
             |i,j \rangle_{3}, |i,j \rangle_{4} \bigr\} ,
\end{align}
where the subscript specifies the four sites.
The Hamiltonian is given by
$H = H_{\rm intra}+H_{\rm inter}$ with~\cite{benalcazar1,benalcazar2}
\begin{align}
   H_{\rm intra}
 & = \sum_{i,j} |i,j \rangle h_{\rm intra} \langle i,j| ,
         \\
   H_{\rm inter}
 & = \sum_{i,j} \bigl\{ |i+1,j \rangle h_{x}^{\lambda} \langle i,j|
                        + {\rm h.c.} \bigr\}
        \\
 & + \sum_{i,j} \bigl\{ |i,j+1 \rangle h_{y}^{\lambda} \langle i,j|
                        + {\rm h.c.} \bigr\}
\end{align}
with
\begin{align}
   h_{\rm intra}
 & = \left[ 
       \begin{array}{cccc}
         E & 0 & \gamma_{x} & \gamma_{y} \\
         0 & E & -\gamma_{y} & \gamma_{x} \\
         \gamma_{x} & -\gamma_{y} & E & 0 \\
         \gamma_{y} & \gamma_{x} & 0 & E \\
       \end{array}
     \right] ,
               \\
   h_{x}^{\lambda}
 & = \left[ 
       \begin{array}{cccc}
         0 & 0 & 0 & 0 \\
         0 & 0 & 0 & \lambda_{x} \\
         \lambda_{x} & 0 & 0 & 0 \\
         0 & 0 & 0 & 0 \\
       \end{array}
     \right] ,
               \\
   h_{y}^{\lambda}
 & = \left[ 
       \begin{array}{cccc}
         0 & 0 & 0 & 0 \\
         0 & 0 & -\lambda_{y} & 0 \\
         0 & 0 & 0 & 0 \\
         \lambda_{y} & 0 & 0 & 0 \\
       \end{array}
     \right] .
\end{align}
As noted in Sect.~5, this model possesses chiral symmetry.
The reference energy $E$, which coincides with the energy of corner states,
is introduced for clarity of our argument and is set equal to zero later.
The system described by the Hamiltonian is topologically nontrivial
under the conditions of~\cite{benalcazar1,benalcazar2}
\begin{align}
      \label{eq:TL}
    -\lambda_{x} < \gamma_{x} < \lambda_{x} ,
         \hspace{4mm}
    -\lambda_{y} < \gamma_{y} < \lambda_{y} ,
\end{align}
where $\lambda_{x}$ and $\lambda_{y}$ are assumed to be positive
without loss of generality.
The model can be regarded as a two-dimensional extension of
the Su--Schrieffer--Heeger model.~\cite{su}

For later convenience, we rewrite the Hamiltonian in terms of dual cells
that are defined by shifting the original unit cells as shown in Fig.~1(b).
Each dual cell is characterized by indices $l$ and $m$ respectively
specifying its location in the $x$- and $y$-directions.
The four-component state vector for the $(l,m)$th dual cell
is expressed as
\begin{align}
  ||l,m \rangle
  =  \bigl\{ ||l,m \rangle_{1}, ||l,m \rangle_{2},
             ||l,m \rangle_{3}, ||l,m \rangle_{4} \bigr\} .
\end{align}
The Hamiltonian is rewritten as
$H = H_{0}+H_{1}$ with
\begin{align}
   H_{0}
 & = \sum_{l,m} ||l,m \rangle h_{\rm 0} \langle l,m|| ,
         \\
   H_{1}
 & = \sum_{l,m}
     \bigl\{ ||l+1,m \rangle h_{x}^{\gamma} \langle l,m|| + {\rm h.c.} \bigr\}
        \\
 & + \sum_{l,m}
     \bigl\{ ||l,m+1 \rangle h_{y}^{\gamma} \langle l,m|| + {\rm h.c.} \bigr\}
\end{align}
with
\begin{align}
   h_{0}
 & = \left[ 
       \begin{array}{cccc}
         E & 0 & \lambda_{x} & \lambda_{y} \\
         0 & E & -\lambda_{y} & \lambda_{x} \\
         \lambda_{x} & -\lambda_{y} & E & 0 \\
         \lambda_{y} & \lambda_{x} & 0 & E \\
       \end{array}
     \right] ,
               \\
   h_{x}^{\gamma}
 & = \left[ 
       \begin{array}{cccc}
         0 & 0 & \gamma_{x} & 0 \\
         0 & 0 & 0 & 0 \\
         0 & 0 & 0 & 0 \\
         0 & \gamma_{x} & 0 & 0 \\
       \end{array}
     \right] ,
               \\
   h_{y}^{\gamma}
 & = \left[ 
       \begin{array}{cccc}
         0 & 0 & 0 & \gamma_{y} \\
         0 & 0 & 0 & 0 \\
         0 & -\gamma_{y} & 0 & 0 \\
         0 & 0 & 0 & 0 \\
       \end{array}
     \right] .
\end{align}

Note that not every dual cell possesses four sites.
If the $(l,m)$th dual cell contains a corner or line edge,
the number of sites is smaller than four; thus, we need to modify
the state vector $||l,m \rangle$ according to its structure.
The dual cells with a corner or line edge are classified in twelve ways
(see Fig.~2).
A $270^\circ$ corner lacks one site; if the $\alpha$th site is lacking,
it is referred to as type $\bar{\alpha}$ ($\alpha  = 1$, $2$, $3$, $4$).
A $90^\circ$ corner contains only one site; if the $\alpha$th site is
contained, it is referred to as type $\alpha$ ($\alpha  = 1$, $2$, $3$, $4$).
A line edge contains two sites aligned horizontally or vertically
in the dual cell; if the $\alpha$th and $\beta$th sites are contained,
it is referred to as type $\alpha\beta$
($\alpha\beta = 13$, $14$, $23$, $24$).
\begin{figure}[btp]
\begin{center}
\includegraphics[height=4.5cm]{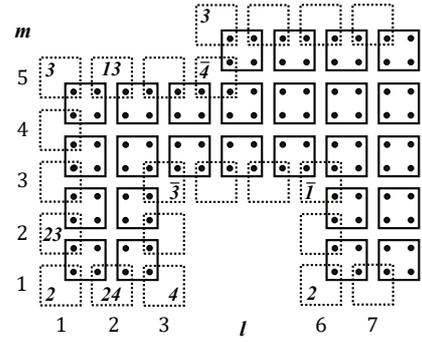}
\end{center}
\caption{
Classification of corners and line edges, where the italic number
in a dotted square denotes the type of corner or line edge contained
in the corresponding dual cell.
}
\end{figure}

If the $(l,m)$th dual cell contains the $270^\circ$ corner
of type $\bar{\alpha}$, we eliminate $||l,m \rangle_{\alpha}$
in $||l,m \rangle$
and represent the resulting state vector as $||l,m \rangle^{\bar{\alpha}}$.
We also represent the corresponding component of $H_{0}$
as $H_{0}^{\bar{\alpha}}$.
In the case of type $\bar{1}$ located at $(l,m)=(6,3)$ in Fig.~2, 
\begin{align}
      \label{eq:H0-bar{1}}
  H_{0}^{\bar{1}}
  = ||l,m \rangle^{\bar{1}}
      \left[ 
         \begin{array}{ccc}
           E & -\lambda_{y} & \lambda_{x} \\
           -\lambda_{y} & E & 0 \\
           \lambda_{x} & 0 & E \\
         \end{array}
       \right] 
    {}^{\bar{1}}\langle l,m||
\end{align}
with
\begin{align}
  ||l,m \rangle^{\bar{1}}
   =  \bigl\{ ||l,m \rangle_{2}, ||l,m \rangle_{3}, ||l,m \rangle_{4} \bigr\} .
\end{align}
If the $(l,m)$th dual cell contains the $90^\circ$ corner of type $\alpha$,
we retain only $||l,m \rangle_{\alpha}$ in $||l,m \rangle$ 
and represent the resulting state vector as $||l,m \rangle^{\alpha}$.
We also represent the corresponding component of $H_{0}$
as $H_{0}^{\alpha}$.
In the case of type $2$ located at $(l,m)=(1,1)$ in Fig.~2,
\begin{align}
      \label{eq:H0-2}
  H_{0}^{2}
  = ||l,m \rangle^{2}\left[ E \right]\, {}^{2}\langle l,m||
\end{align}
with $||l,m \rangle^{2} = ||l,m \rangle_{2}$.
If the $(l,m)$th dual cell contains the line edge of type $\alpha\beta$,
we retain $||l,m \rangle_{\alpha}$ and $||l,m \rangle_{\beta}$
in $||l,m \rangle$
and represent the resulting state vector as
$||l,m \rangle^{\alpha\beta}$.
We also represent the corresponding component of $H_{0}$
as $H_{0}^{\alpha \beta}$.
In the case of type $24$ located at $(l,m)=(2,1)$ in Fig.~2,
\begin{align}
      \label{eq:H0-24}
 H_{0}^{24}
  = ||l,m \rangle^{24}
      \left[ 
        \begin{array}{cc}
          E & \lambda_{x} \\
          \lambda_{x} & E \\
        \end{array}
      \right] 
    {}^{24}\langle l,m||
\end{align}
with 
\begin{align}
  ||l,m \rangle^{24}
  =  \bigl\{ ||l,m \rangle_{2}, ||l,m \rangle_{4} \bigr\} .
\end{align}

We focus on corner states in the limit of no electron transfer
in the unit cell (i.e., $\gamma_{x} = \gamma_{y} = 0$),
in which the system is topologically nontrivial
in accordance with Eq.~(\ref{eq:TL}).
In this limit, each dual cell is completely disconnected from
neighboring ones, so that electron states in a dual cell are
fully described by the corresponding component of $H_{0}$.
Hence, we can easily obtain the wavefunction of various corner states,
the energy of which is equal to $E$.
If the $(l,m)$th dual cell contains the $90^\circ$ corner of type $2$,
the component of $H_{0}$ for this cell is given in Eq.~(\ref{eq:H0-2}).
The corner state is given by
\begin{align}
      \label{eq:phi-0_2}
  |\psi_{2} \rangle_{l,m} = ||l,m \rangle_{2} ,
\end{align}
indicating that it has a finite amplitude only at the second site in this cell.
If the $(l,m)$th dual cell contains the $270^\circ$ corner of
type $\bar{1}$, the wavefunction is obtained by diagonalizing 
the component of $H_{0}$ for this cell given in Eq.~(\ref{eq:H0-bar{1}}).
The corner state is
\begin{align}
      \label{eq:phi-0_bar1}
  |\psi_{\bar{1}} \rangle_{l,m}
   = \frac{\lambda_{x}}{\sqrt{\Theta}} ||l,m \rangle_{3}
   + \frac{\lambda_{y}}{\sqrt{\Theta}} ||l,m \rangle_{4}
\end{align}
with
\begin{align}
  \Theta = \lambda_{x}^{2}+\lambda_{y}^{2} .
\end{align}
This indicates that it has a finite amplitude only
at the third and fourth sites in the dual cell.
If the corner is of type $\bar{3}$, the corner state is given by
\begin{align}
  |\psi_{\bar{3}} \rangle_{l,m}
   = \frac{\lambda_{x}}{\sqrt{\Theta}} ||l,m \rangle_{1}
   - \frac{\lambda_{y}}{\sqrt{\Theta}} ||l,m \rangle_{2} .
\end{align}

\section{Formulation and Simple Application}

Setting $E = 0$, we hereafter consider
only a zero-energy edge-localized state in semi-infinite systems.
The case of a finite system is briefly considered in Sect.~5.
We assume that the edge structure of a system consists of $N$ corners,
each of which is the $90^{\circ}$ or $270^{\circ}$ one,
and that the corners are sequentially numbered along the edge.
Let us introduce index $\zeta_{q}$ to classify the type of
$q$th corner ($1 \le q \le N$) in the dual cell at $(l_{q},m_{q})$:
$\zeta_{q} = \alpha$ if the corner is the $90^{\circ}$ one of
type $\alpha$ and $\zeta_{q} = \bar{\alpha}$ if the corner is
the $270^{\circ}$ one of type $\bar{\alpha}$.

Let $|\psi \rangle$ be a zero-energy eigenfunction of $H_{0}$
satisfying $H_{0}|\psi \rangle = 0$.
This is equivalent to saying that $|\psi \rangle$ is an eigenfunction of $H$
in the limit of $\gamma_{x} = \gamma_{y} = 0$.
In this limit, the $q$th corner hosts a zero-energy corner state that is
described by $|\psi_{\zeta_{q}}\rangle_{l_{q},m_{q}}$.
Hence, $|\psi \rangle$ is generally written as a superposition of
the zero-energy corner states:
\begin{align}
       \label{eq:multi-corners}
  |\psi \rangle = \sum_{q = 1}^{N} f_{q}
                  |\psi_{\zeta_{q}}\rangle_{l_{q},m_{q}} ,
\end{align}
where $\{f_{q}\}$ is a set of arbitrary constants.
Starting from $|\psi \rangle$, we attempt to describe
a zero-energy edge localized state in the case of
$\gamma_{x} \neq 0$ and $\gamma_{y} \neq 0$.
Our attention is focused on the question of
whether an edge localized state can appear at zero energy
even when $\gamma_{x} \neq 0$ and $\gamma_{y} \neq 0$.
To answer this, we use the fact that a zero-energy eigenfunction
$|\Psi \rangle$ of $H = H_{0}+H_{1}$ is formally expressed as
\begin{align}
      \label{eq:WO-rep}
  |\Psi \rangle
   = \sum_{p = 0}^{\infty}\left(\frac{1}{-H_{0}+i\delta}H_{1}\right)^{p}
     |\psi \rangle ,
\end{align}
where $\delta$ is an infinitesimal.
If this series converges in the limit of $\delta \to 0$
for a given $\{f_{q}\}$, the resulting function satisfies
the eigenvalue equation of $H|\Psi \rangle = 0$.
That is, $|\Psi \rangle$ is the eigenfunction of $H$
representing a zero-energy edge localized state.
Note that $(-H_{0}+i\delta)^{-1}H_{1}$ induces various hopping processes
of an electron as shown in Fig.~3.
\begin{figure}[btp]
\begin{center}
\includegraphics[height=3.4cm]{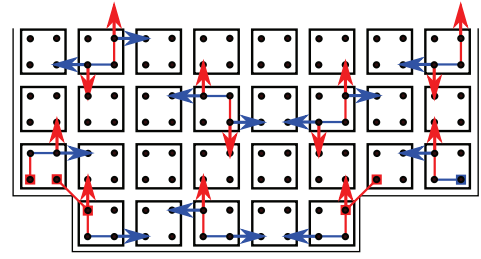}
\end{center}
\caption{
(Color online) Hopping processes involved in $(-H_{0}+i\delta)^{-1}H_{1}$,
where filled squares indicate the appearance of a singular term
with $(i\delta)^{-1}$.
}
\end{figure}

We clarify the singularity of $(-H_{0}+i\delta)^{-1}$,
which is directly related to the zero-energy corner states in the limit of
$\gamma_{x} = \gamma_{y} = 0$, using the following expression:
\begin{align}
     \label{eq:def-inv}
   (-H_{0}+i\delta)^{-1}
 & = \sum_{l,m} ||l,m \rangle \Lambda \langle l,m|| ,
\end{align}
where $\Lambda = (-h_{\rm 0}+i\delta)^{-1}$.
Remember that the number of sites in a dual cell is smaller than four
if the cell contains a corner or line edge.
Therefore, the explicit form of $\Lambda$ depends on
whether the dual cell contains a corner or line edge.
In the absence of both a corner and a line edge, $\Lambda$ is given by
\begin{align}
   \Lambda
   = \frac{1}{\Theta}
     \left[ 
       \begin{array}{cccc}
         0 & 0 & -\lambda_{x} & -\lambda_{y} \\
         0 & 0 & \lambda_{y} & -\lambda_{x} \\
         -\lambda_{x} & \lambda_{y} & 0 & 0 \\
         -\lambda_{y} & -\lambda_{x} & 0 & 0 \\
       \end{array}
     \right] .
\end{align}
If the $(l,m)$th dual cell contains a line edge of type $\alpha \beta$,
we need to replace $||l,m \rangle$ with $||l,m \rangle^{\alpha\beta}$
and $\Lambda$ with $\Lambda^{\alpha\beta}$ in Eq.~(\ref{eq:def-inv}),
where $\Lambda^{\alpha\beta}$ takes the effect of the line edge into account.
The four $\Lambda^{\alpha\beta}$ are given by
\begin{align}
 & \Lambda^{13} = \Lambda^{24}
   = \frac{-1}{\lambda_{x}}
     \left[ 
       \begin{array}{cc}
         0 & 1 \\
         1 & 0 \\
       \end{array}
     \right] ,
       \\
 & \Lambda^{14} = - \Lambda^{23}
   = \frac{-1}{\lambda_{y}}
     \left[ 
       \begin{array}{cccc}
         0 & 1 \\
         1 & 0 \\
       \end{array}
     \right] .
\end{align}
In the above expressions, $i\delta$ can be safely ignored.
If the $(l,m)$th dual cell contains
the $270^\circ$ corner of type $\bar{\alpha}$,
we need to replace $||l,m \rangle$ with $||l,m \rangle^{\bar{\alpha}}$
and $\Lambda$ with $\Lambda^{\bar{\alpha}}$, where $\Lambda^{\bar{\alpha}}$
takes the effect of the $270^\circ$ corner into account.
The four $\Lambda^{\bar{\alpha}}$ are given by
\begin{align}
      \label{eq:L-bar1}
 & \Lambda^{\bar{1}}
   = \frac{1}{\Theta}
     \left[ 
       \begin{array}{ccc}
         0 & \lambda_{y} & -\lambda_{x} \\
         \lambda_{y} & \frac{\lambda_{x}^{2}}{i\delta}
                  & \frac{\lambda_{x}\lambda_{y}}{i\delta} \\
         -\lambda_{x} & \frac{\lambda_{x}\lambda_{y}}{i\delta}
                  & \frac{\lambda_{y}^{2}}{i\delta} \\
       \end{array}
     \right] ,
          \\
 & \Lambda^{\bar{2}}
   = \frac{1}{\Theta}
     \left[ 
       \begin{array}{ccc}
         0 & -\lambda_{x} & -\lambda_{y} \\
         -\lambda_{x} & \frac{\lambda_{y}^{2}}{i\delta}
                  & -\frac{\lambda_{x}\lambda_{y}}{i\delta} \\
         -\lambda_{y} & -\frac{\lambda_{x}\lambda_{y}}{i\delta}
                  & \frac{\lambda_{x}^{2}}{i\delta} \\
       \end{array}
     \right] ,
          \\
 & \Lambda^{\bar{3}}
   = \frac{1}{\Theta}
     \left[ 
       \begin{array}{ccc}
         \frac{\lambda_{x}^{2}}{i\delta}
                  & -\frac{\lambda_{x}\lambda_{y}}{i\delta} & -\lambda_{y} \\
         -\frac{\lambda_{x}\lambda_{y}}{i\delta}
                  & \frac{\lambda_{y}^{2}}{i\delta} & -\lambda_{x} \\
         -\lambda_{y} & -\lambda_{x} & 0 \\
       \end{array}
     \right] ,
          \\
 & \Lambda^{\bar{4}}
   = \frac{1}{\Theta}
     \left[ 
       \begin{array}{ccc}
         \frac{\lambda_{y}^{2}}{i\delta}
                  & \frac{\lambda_{x}\lambda_{y}}{i\delta} & -\lambda_{x} \\
         \frac{\lambda_{x}\lambda_{y}}{i\delta}
                  & \frac{\lambda_{x}^{2}}{i\delta} & \lambda_{y} \\
         -\lambda_{x} & \lambda_{y} & 0 \\
       \end{array}
     \right] .
\end{align}
Finally, if the $(l,m)$th dual cell contains
the $90^\circ$ corner of type $\alpha$,
we need to replace $||l,m \rangle$ with $||l,m \rangle^{\alpha}$
and $\Lambda$ with
\begin{align}
      \label{eq:L-alpha}
   \Lambda^{\alpha} = \left[ \frac{1}{i\delta} \right] .
\end{align}
Note that $\Lambda^{\alpha}$ and $\Lambda^{\bar{\alpha}}$
are singular at the sites
where the corresponding zero-energy corner state has a finite amplitude.
For example, $\Lambda^{2}$ is singular at the second site
[see Eq.~(\ref{eq:phi-0_2})] and $\Lambda^{\bar{1}}$ is singular at
the third and fourth sites [see Eq.~(\ref{eq:phi-0_bar1})].
These sites are referred to as singular sites.
In addition, a dual cell with one or two singular sites
is referred to as a singular cell.
\begin{figure}[btp]
\begin{center}
\includegraphics[height=3.0cm]{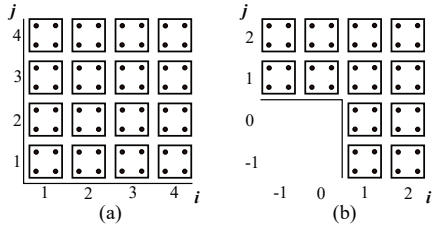}
\end{center}
\caption{
One-corner case analyzed in the text:
(a) semi-infinite system with a $90^{\circ}$ corner of type $2$ and
(b) semi-infinite system with a $270^{\circ}$ corner of type $\bar{1}$.
}
\end{figure}

We apply the method to the case of $N = 1$,
in which the semi-infinite system includes
only one $90^{\circ}$ or $270^{\circ}$ corner.
Since no singular terms appear in $|\Psi \rangle$ in this case,
we are allowed to ignore the singularity involved in $(-H_{0}+i\delta)^{-1}$.
In the remainder of this section, we use the unit cell representation.
Let us determine the wavefunction of a zero-energy corner state
in the system with the $90^{\circ}$ corner of type $2$ [see Fig.~4(a)],
for which $|\psi \rangle$ in Eq.~(\ref{eq:WO-rep}) is identified as
\begin{align}
  |\psi \rangle = |1,1 \rangle_{2} .
\end{align}
Substituting this into Eq.~(\ref{eq:WO-rep}),
we perturbatively obtain $|\Psi \rangle$ term by term and find
\begin{align}
  |\Psi \rangle
   = c \sum_{i,j=1}^{\infty}
     \left(-\frac{\gamma_{x}}{\lambda_{x}}\right)^{i-1}
     \left(-\frac{\gamma_{y}}{\lambda_{y}}\right)^{j-1}
     |i,j \rangle_{2} ,
\end{align}
where $c$ is the normalization constant.

We next determine the wavefunction of a zero-energy corner state
in the system with the $270^{\circ}$ corner of type $\bar{1}$
[see Fig.~4(b)], for which $|\psi \rangle$ is identified as
\begin{align}
  |\psi \rangle
   = \frac{\lambda_{x}}{\sqrt{\Theta}} |1,0 \rangle_{3}
   + \frac{\lambda_{y}}{\sqrt{\Theta}} |0,1 \rangle_{4} .
\end{align}
Substituting this into Eq.~(\ref{eq:WO-rep}), we observe that
$|\Psi \rangle$ is represented in the from of
\begin{equation}
       \label{eq:phi_c-d}
  |\Psi \rangle
   = \sum_{i,j= -\infty}^{\infty}
     \bigl( c_{i,j} |i,j \rangle_{3} + d_{i,j} |i,j \rangle_{4} \bigr) ,
\end{equation}
where $c_{i,j} = d_{i,j} = 0$ if $i \le 0$ and $j \le 0$.
As the perturbative determination of $|\Psi \rangle$ is not easy in this case,
we numerically calculate $c_{i,j}$ and $d_{i,j}$.
Figure~5 shows the resulting $c_{i,j}$ and $d_{i,j}$ multiplied by
$(-1)^{i+j-1}$ [i.e., $(-1)^{i+j-1}c_{i,j}$ and $(-1)^{i+j-1}d_{i,j}$].
Here and hereafter, the parameters are set as $\gamma_{x}/\lambda_{x} = 0.8$,
$\gamma_{y}/\lambda_{y} = 0.6$, and $\lambda_{y}/\lambda_{x} = 1.2$
in numerical calculations.
\begin{figure}[btp]
\begin{center}
\includegraphics[height=4.6cm]{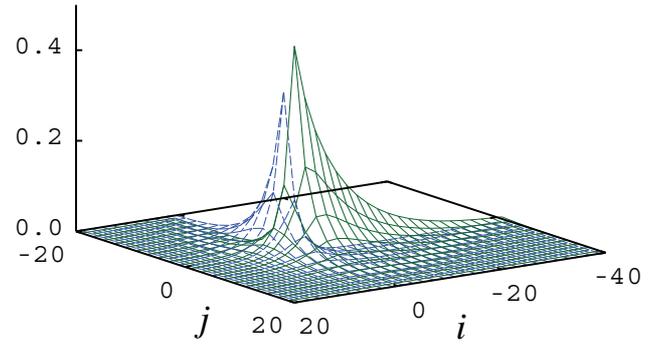}
\end{center}
\caption{
(Color online) Zero-energy wavefunction for the semi-infinite system
with a $270^{\circ}$ corner as shown in Fig.~4(b),
where dashed lines (blue) and solid lines (green) respectively represent
$(-1)^{i+j-1}c_{i,j}$ and $(-1)^{i+j-1}d_{i,j}$.
}
\end{figure}

\section{Application to Multicorner Cases}

Let us consider the semi-infinite system with $N$ corners characterized by
$\{\zeta_{q}\}$ and $\{(l_{q},m_{q})\}$ with $1 \le q \le N$
[see Figs.~6(a) and 6(b) as examples].
The wavefunction for a zero-energy edge localized state is obtained
if $|\Psi \rangle$ defined in Eq.~(\ref{eq:WO-rep}) converges
in the limit of $\delta \to 0$ for a given set of $\{f_{q}\}$.
Here, we show that a zero-energy state exists
only when $N$ is an odd integer.

To do so, it is convenient to decompose $|\psi \rangle$ as
$|\psi \rangle = |\psi \rangle_{\rm odd}+|\psi \rangle_{\rm even}$ with
\begin{align}
    |\psi \rangle_{\rm odd}
 & = \sum_{n = 1}^{n_{\rm odd}}
      f_{2n-1}|\psi_{\zeta_{2n-1}}\rangle_{l_{2n-1},m_{2n-1}} ,
             \\
    |\psi \rangle_{\rm even}
 & = \sum_{n = 1}^{n_{\rm even}}
      f_{2n}|\psi_{\zeta_{2n}}\rangle_{l_{2n},m_{2n}} .
\end{align}
In the odd case of $N = 2n_{\rm c}-1$
with $n_{\rm c}$ being a positive integer,
$n_{\rm odd} = n_{\rm c}$ and $n_{\rm even} = n_{\rm c}-1$,
whereas $n_{\rm odd} = n_{\rm even} = n_{\rm c}$
in the even case of $N = 2n_{\rm c}$.
We next define $|\Psi \rangle_{\rm odd}$ and $|\Psi \rangle_{\rm even}$ as
\begin{align}
   |\Psi \rangle_{\rm odd}
 & = \sum_{p = 0}^{\infty}
     \left(\frac{1}{-H_{0}+i\delta}H_{1}\right)^{p}
     |\psi \rangle_{\rm odd} ,
            \\
   |\Psi \rangle_{\rm even}
 & = \sum_{p = 0}^{\infty}
     \left(\frac{1}{-H_{0}+i\delta}H_{1}\right)^{p}
     |\psi \rangle_{\rm even} .
\end{align}
It is important to point out that $|\Psi \rangle_{\rm odd}$ and
$|\Psi \rangle_{\rm even}$ possess singular terms in different sets
of singular cells: the singular terms of $|\Psi \rangle_{\rm odd}$ appear
in the singular cells, related to $|\psi \rangle_{\rm even}$,
at $(l_{2n'},m_{2n'})$ with $1 \le n' \le n_{\rm even}$
and those of $|\Psi \rangle_{\rm even}$ appear in the singular cells,
related to $|\psi \rangle_{\rm odd}$,
at $(l_{2n'-1},m_{2n'-1})$ with $1 \le n' \le n_{\rm odd}$.
This statement is justified by operating on $|\psi \rangle_{\rm odd}$ and
$|\psi \rangle_{\rm even}$ with $(-H_{0}+i\delta)^{-1}H_{1}$
in a successive manner.
Two examples are given in the last part of this section.
When $N = 2n_{\rm c}-1$, the singular terms of $|\Psi \rangle_{\rm odd}$
appear in the $n_{\rm c}-1$ singular cells, whereas $|\Psi \rangle_{\rm odd}$
consists of the $n_{\rm c}$ undetermined coefficients.
Hence, the singular terms of $|\Psi \rangle_{\rm odd}$ can be canceled out
by determining $\{f_{2n-1}\}$ in an appropriate manner.
Note that only one solution exists.
In contrast, the singular terms of $|\Psi \rangle_{\rm even}$ appear
in the $n_{\rm c}$ singular cells, whereas $|\Psi \rangle_{\rm even}$
consists of the $n_{\rm c}-1$ undetermined coefficients.
Hence, the singular terms of $|\Psi \rangle_{\rm even}$ cannot be
canceled out, indicating that $|\Psi \rangle$ converges
only when $|\Psi \rangle_{\rm even} = 0$
(i.e., $f_{2n} = 0$ for $1 \le n \le n_{\rm c}-1$).
That is, we can obtain a converged solution at zero energy
by determining $\{f_{2n-1}\}$
under the condition of $f_{2n} = 0$ for $1 \le n \le n_{\rm c}-1$.
A similar consideration shows that
$|\Psi \rangle$ cannot converge if $N = 2n_{\rm c}$.
In summary, a zero-energy state exists
only in the odd case with $N = 2n_{\rm c}-1$.
\begin{figure}[btp]
\begin{center}
\includegraphics[height=8.5cm, angle=90]{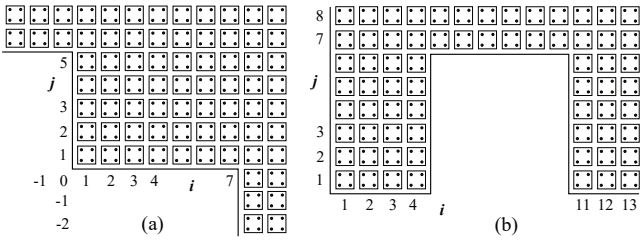}
\end{center}
\caption{
Multicorner cases analyzed in the text:
(a) semi-infinite system with three corners
and (b) semi-infinite system with five corners.
}
\end{figure}

Now, we present a simple method of determining $\{f_{q}\}$
for the case of $N = 2n_{\rm c} - 1$
under the condition of $f_{2n} = 0$ for $1 \le n \le n_{\rm c}-1$.
This method, referred to as the simplified method, is based on the fact
that the convergence of $|\Psi \rangle$ in the limit of $\delta \to 0$ is
equivalent to requiring that $|\Psi \rangle$ satisfies
the eigenvalue equation in the singular cells (see Appendix).
As the first step, we numerically obtain
$|\Psi \rangle_{2n-1}^{\rm reg}$ defined by
\begin{align}
     \label{eq:def-Psi_2n-1}
  |\Psi \rangle_{2n-1}^{\rm reg}
   = \sum_{p = 0}^{\infty}
     \left(\left[\frac{1}{-H_{0}+i\delta}\right]_{\rm reg}H_{1}\right)^{p}
     |\psi_{\zeta_{2n-1}}\rangle_{l_{2n-1},m_{2n-1}} ,
\end{align}
where $[\cdots ]_{\rm reg}$ indicates that the singular terms with respect to
$(i\delta)^{-1}$ are completely removed.
The general solution is written as
\begin{align}
  |\Psi \rangle
  = \sum_{n = 1}^{n_{\rm c}} f_{2n-1}|\Psi \rangle_{2n-1}^{\rm reg} .
\end{align}
This does not satisfy the eigenvalue equation in the singular cells
at $(l_{2n'},m_{2n'})$ with $1 \le n' \le n_{\rm c}-1$  (see Appendix).
Using the numerical result of $|\Psi \rangle_{2n-1}^{\rm reg}$
with $1 \le n \le n_{\rm c}$,
we can determine $\{f_{2n-1}\}$ such that $|\Psi \rangle$ satisfies
the eigenvalue equation in the singular cells denoted above.
The solution is the wavefunction of a zero-energy edge localized state.

\begin{figure}[btp]
\begin{center}
\includegraphics[height=4.6cm]{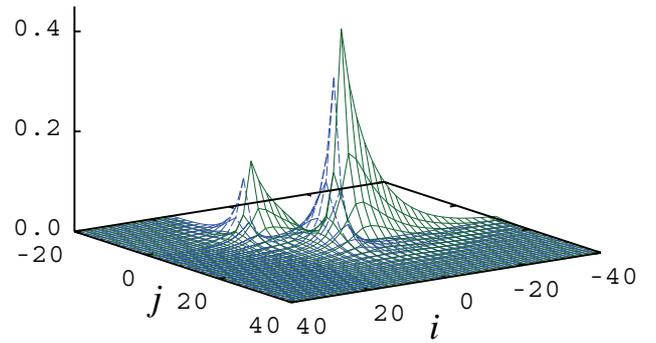}
\end{center}
\caption{
(Color online) Zero-energy wavefunction for the semi-infinite system
with three corners as shown in Fig.~6(a),
where dashed lines (blue) and solid lines (green) respectively represent
$(-1)^{i+j}c_{i,j}$ and $(-1)^{i+j}d_{i,j}$.
}
\end{figure}
Let us apply the simplified method to the three-corner case
as shown in Fig.~6(a).
In the unit cell representation, $|\psi \rangle_{\rm odd}$ and
$|\psi \rangle_{\rm even}$ are given by
\begin{align}
        \label{eq:tri-corners-odd}
  |\psi \rangle_{\rm odd}
 & =  f_{1}\left( |1,5 \rangle_{3}
                 + \frac{\lambda_{y}}{\lambda_{x}}|0,6 \rangle_{4}
           \right)
                \nonumber \\
 & \hspace{20mm}
   + f_{3}\left( |8,0 \rangle_{3}
                 + \frac{\lambda_{y}}{\lambda_{x}}|7,1 \rangle_{4}
           \right) ,
                \\
        \label{eq:tri-corners-even}
   |\psi \rangle_{\rm even} & = f_{2}|1,1 \rangle_{2}.
\end{align}
If $(-H_{0}+i\delta)^{-1}H_{1}$ successively operates on
$|\psi \rangle_{\rm even}$, the singular terms appear
at the third and fourth sites
in the unit cells at $(i,j) = (1,5)$ and $(0,6)$
and those in the unit cells at $(i,j) = (8,0)$ and $(7,1)$, respectively.
These singular terms do not disappear as long as $f_{2} \neq 0$,
indicating that $f_{2} = 0$ is a necessary condition
for obtaining a zero-energy wavefunction.
If $(-H_{0}+i\delta)^{-1}H_{1}$ successively operates on
$|\psi \rangle_{\rm odd}$, singular terms appear
at the second site in the unit cell at $(i,j) = (1,1)$.
These terms can be canceled out
if $f_{1}$ and $f_{3}$ are determined in an appropriate manner.
We numerically obtain $|\Psi \rangle_{1}^{\rm reg}$
and $|\Psi \rangle_{3}^{\rm reg}$ defined in Eq.~(\ref{eq:def-Psi_2n-1}),
in terms of which the general solution is given by
\begin{align}
  |\Psi \rangle = f_{1}|\Psi \rangle_{1}^{\rm reg}
                + f_{3}|\Psi \rangle_{3}^{\rm reg} .
\end{align}
According to the simplified method, the wavefunction at zero energy is
obtained by determining $f_{1}$ and $f_{3}$
such that $|\Psi \rangle$ satisfies the eigenvalue equation
at the second site in the unit cell at $(i,j) = (1,1)$.
In this case,
$|\Psi \rangle$ is expressed in the form of Eq.~(\ref{eq:phi_c-d}).
Figure~7 shows the resulting $(-1)^{i+j}c_{i,j}$ and $(-1)^{i+j}d_{i,j}$.
\begin{figure}[btp]
\begin{center}
\includegraphics[height=4.6cm]{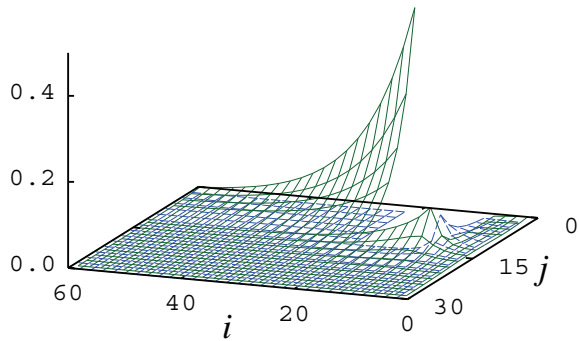}
\end{center}
\caption{
(Color online) Zero-energy wavefunction for the semi-infinite system
with five corners as shown in Fig.~6(b),
where dashed lines (blue) and solid lines (green) respectively represent
$(-1)^{i+j-1}a_{i,j}$ and $(-1)^{i+j}b_{i,j}$.
}
\end{figure}

Let us turn to the five-corner case as shown in Fig.~6(b).
In the unit cell representation, $|\psi \rangle_{\rm odd}$ and
$|\psi \rangle_{\rm even}$ are written as
\begin{align}
        \label{eq:penta-corners-odd}
  |\psi \rangle_{\rm odd}
 & =  f_{1}|1,1 \rangle_{2}
    + f_{3}\left( |4,6 \rangle_{1}
                 - \frac{\lambda_{y}}{\lambda_{x}}|5,7 \rangle_{2}
           \right) + f_{5}|11,1 \rangle_{2} ,
                \\
        \label{eq:penta-corners-even}
 |\psi \rangle_{\rm even}
 & = f_{2}|4,1 \rangle_{4}
   + f_{4}\left( |11,6 \rangle_{3}
                 + \frac{\lambda_{y}}{\lambda_{x}}|10,7 \rangle_{4}
           \right) .
\end{align}
If $(-H_{0}+i\delta)^{-1}H_{1}$ successively operates on
$|\psi \rangle_{\rm even}$, the singular terms appear at the second sites
in the unit cells at $(i,j) = (1,1)$ and $(11,1)$
and the first and second sites in the unit cells
at $(i,j) = (4,6)$ and $(5,7)$, respectively.
These singular terms disappear only when $f_{2} = f_{4} = 0$.
If $(-H_{0}+i\delta)^{-1}H_{1}$ successively operates on
$|\psi \rangle_{\rm odd}$, the singular terms appear
at the fourth site in the unit cell at $(i,j) = (4,1)$
and the third and fourth sites in the unit cells
at $(i,j) = (11,6)$ and $(10,7)$, respectively.
These singular terms can be canceled out if $f_{1}$, $f_{3}$, and $f_{5}$
are determined in an appropriate manner.
We numerically obtain $|\Psi \rangle_{1}^{\rm reg}$,
$|\Psi \rangle_{3}^{\rm reg}$, and $|\Psi \rangle_{5}^{\rm reg}$,
in terms of which the general solution is given by
\begin{align}
  |\Psi \rangle
  = f_{1}|\Psi \rangle_{1}^{\rm reg}
  + f_{3}|\Psi \rangle_{3}^{\rm reg} + f_{5}|\Psi \rangle_{5}^{\rm reg} .
\end{align}
According to the simplified method, the wavefunction at zero energy is
obtained by determining $f_{1}$, $f_{3}$, and $f_{5}$
such that $|\Psi \rangle$ satisfies the eigenvalue equation
at the fourth site in the unit cell at $(i,j) = (4,1)$
and the third and fourth sites in the unit cells at $(i,j) = (11,6)$ and
$(10,7)$, respectively [see Eqs.~(\ref{eq:eve1}) and (\ref{eq:eve2})].
In this case, $|\Psi \rangle$ is expressed in the form of
\begin{equation}
       \label{eq:phi_a-b}
  |\Psi \rangle
   = \sum_{i,j= 1}^{\infty}
     \bigl( a_{i,j} |i,j \rangle_{1} + b_{i,j} |i,j \rangle_{2} \bigr) .
\end{equation}
Figure~8 shows the resulting $(-1)^{i+j-1}a_{i,j}$ and $(-1)^{i+j}b_{i,j}$.

\section{Summary and Discussion}

We presented a numerical method of determining the wavefunction of
a zero-energy state localized near an arbitrary edge
in quadrupole topological insulators.
Applying it to several cases, we showed that this method is
practically useful in determining the wavefunction of a zero-energy state.

As a byproduct of the method, we found that one localized state appears
at $E = 0$ if the edge consists of an odd number of corners,
whereas the energy of localized states inevitably deviates from $E = 0$
if the edge includes an even number of corners.
This feature is explained on the basis of the chiral symmetry
of the model Hamiltonian:
\begin{align}
  \Gamma^{-1} H \Gamma = - H
\end{align}
with $\Gamma = \sigma_{z} \otimes \sigma_{0}$,
where $\sigma_{z}$ and $\sigma_{0}$ are respectively the $z$-component of
Pauli matrices and the $2\times2$ unit matrix.
This symmetry ensures that except at zero energy, edge localized states
appear inside the band gap in pairs: if one state has energy $\epsilon$,
the other has energy $-\epsilon$.
Note that if the edge consists of $N$ corners,
the number of zero-energy corner states is equal to $N$
in the limit of $\gamma_{x} = \gamma_{y} = 0$,
indicating that $N$ midgap states appear
when $\gamma_{x} \neq 0$ and $\gamma_{y} \neq 0$.
Combining this with the paired nature of edge localized states,
we conclude that one state must appear at $E = 0$ if $N$ is an odd integer,
whereas no state appears at $E = 0$ if $N$ is an even integer.
This is consistent with the observation given in Sect.~4.

As explained above, our method relies on the chiral symmetry ensuring
the existence of a zero-energy state unless $N$ is an even integer.
In other words, it is not applicable to a model system in which the energy
of a corner state varies if a relevant parameter of the system is changed.
Our method requires the chiral symmetry, or an alternative condition,
in addition to reflection symmetry~\cite{benalcazar1,benalcazar2}
that guarantees the presence of a corner state.

Let us finally consider the applicability of our method to
a finite system with multiple $90^{\circ}$ and $270^{\circ}$ corners.
Strictly speaking, the method cannot be applied to a finite system
as it inevitably includes an even number of corners.
That is, an edge localized state in a finite system has a nonzero energy.
However, if the energy is very close to zero, we can apply the method
to obtain an approximate wavefunction.
For example, if the entire edge of a finite system
can be decomposed into several local edge structures
and one of them is sufficiently separated from the others,
we can apply the method to the separated edge structure
if the number of corners in it is an odd integer.

\section*{Acknowledgment}

This work was supported by JSPS KAKENHI Grant Number JP18K03460.

\appendix

\section{}

Considering the five-corner case shown in Fig.~6(b), we show that
the convergence of $|\Psi \rangle$ in the limit of $\delta \to 0$ is
equivalent to requiring that
$|\Psi \rangle$ satisfies the eigenvalue equation in the singular cells.
Hereafter, we use the unit cell representation.

We consider $|\Psi \rangle$ defined in Eq.~(\ref{eq:WO-rep})
with $f_{2} = f_{4} = 0$.
Note that the singular term of $|\Psi \rangle$ appears
at three singular sites: the fourth site in the unit cell
at $(i,j) = (4,1)$ and the third and fourth sites in the unit cells
at $(i,j) = (11,6)$ and $(10,7)$, respectively.
Let us assume that $f_{1}$, $f_{3}$, and $f_{5}$ are appropriately determined
such that $|\Psi \rangle$ converges in the limit of $\delta \to 0$.
Under this assumption, no singular term appears
if $(-H_{0}+i\delta)^{-1}H_{1}$ operates on $|\Psi \rangle$.
Focusing on terms that are directly related to the singular sites
indicated above, we write $|\Psi \rangle$ as
\begin{align}
  |\Psi \rangle
 & = a_{4,1}|4,1 \rangle_{1} + b_{4,1}|4,1 \rangle_{2}
   + a_{10,7}|10,7 \rangle_{1} + b_{10,7}|10,7 \rangle_{2}
               \nonumber \\
 & + a_{11,6}|11,6 \rangle_{1} + b_{11,6}|11,6 \rangle_{2}
   + b_{11,7}|11,7 \rangle_{2} + \cdots .
\end{align}
If $(-H_{0}+i\delta)^{-1}H_{1}$ operates on $|\Psi \rangle$,
the singular terms seemingly appear at the singular sites as
\begin{align}
     \label{eq:sing-result}
 & \left( \gamma_{y}a_{4,1} + \gamma_{x}b_{4,1} \right)
   \frac{|4,1 \rangle_{4}}{i\delta}
           \nonumber \\
 & + \Bigl[ \lambda_{y}\left(\gamma_{y}a_{10,7}+\gamma_{x}b_{10,7}\right)
          + \lambda_{x}\left(\gamma_{x}a_{11,6}-\gamma_{y}b_{11,6}\right)
     \Bigr]
           \nonumber \\
 & \hspace{15mm} \times
     \frac{\lambda_{x}|11,6 \rangle_{3}+\lambda_{y}|10,7 \rangle_{4}}
          {\Theta i\delta} ,
\end{align}
where Eqs.~(\ref{eq:L-bar1}) and (\ref{eq:L-alpha}) are used.
Since $|\Psi \rangle$ has no singular term, the coefficients satisfy
\begin{align}
    \label{eq:eve1}
 & \gamma_{y}a_{4,1} + \gamma_{x}b_{4,1} = 0 ,
               \\
    \label{eq:eve2}
 & \frac{\gamma_{y}a_{10,7}+\gamma_{x}b_{10,7}}{\lambda_{x}}
   + \frac{\gamma_{x}a_{11,6}-\gamma_{y}b_{11,6}}{\lambda_{y}} = 0 .
\end{align}
Equation~(\ref{eq:eve1}) is exactly the eigenvalue equation
at the fourth site in the unit cell at $(i,j) = (4,1)$
in the case of $E = 0$.
Equation~(\ref{eq:eve2}) is equivalent to the eigenvalue equation
at the third and fourth sites in the unit cells
at $(i,j) = (11,6)$ and $(10,7)$, respectively.
Indeed, the eigenvalue equation at these sites is expressed as
\begin{align}
 & \gamma_{y}a_{10,7}+\gamma_{x}b_{10,7} + \lambda_{x}b_{11,7} = 0 ,
        \\
 & \gamma_{x}a_{11,6}-\gamma_{y}b_{11,6} - \lambda_{y}b_{11,7} = 0 ,
\end{align}
which are combined to give Eq.~(\ref{eq:eve2}).

Although this argument is based on a particular case,
the conclusion is general; the vanishing of singular terms is
equivalent to ensuring the eigenvalue equation at the singular sites.


\begin{thebibliography}{99}
\bibitem{comment} We implicitly assume that
the rectangular system is very large.
If this is not the case, the energy $E$ of the corner state
weakly depends on the system size.
\end{thebibliography}
\end{document}